\documentclass[pt11,aps,superscriptaddress,notitlepage]{revtex4-1}%
\usepackage{amsmath}
\usepackage{amssymb}
\usepackage{amsfonts}
\usepackage{revsymb}
\usepackage{graphicx}
\usepackage{color}
\usepackage{hyperref}%
\providecommand{\U}[1]{\protect\rule{.1in}{.1in}}

\def\F{{\mathbb{F}}}
\def\CHSH{{\mathrm{CHSH}}}
\begin{document}

\title{An explicit classical strategy for winning a $\mathrm{CHSH}_{q}$ game}
\author{Matej Pivoluska}

\affiliation{Faculty of Informatics, Masaryk University,  Botanick\'a 68a, 602 00 Brno, Czech Republic}

\author{Martin Plesch}

\affiliation{Faculty of Informatics, Masaryk University,  Botanick\'a 68a, 602 00 Brno, Czech Republic}
\affiliation{Institute of Physics, Slovak Academy of Sciences,
Bratislava, Slovakia}

\begin{abstract}
A $\mathrm{CHSH}_{q}$ game is a generalization of the standard two player $\mathrm{CHSH}$ game, 
having $q$ different input and output options. In contrast to the binary game,
the best classical and quantum winning strategies are not known exactly. In this paper
we provide a constructive classical strategy for winning a $\mathrm{CHSH}_{q}$
game, with $q$ being a prime. Our construction achieves a winning
probability better than $\frac{1}{22}q^{-\frac{2}{3}}$, which is in contrast with the 
previously known constructive strategies achieving only the winning probability of
$O(q^{-1})$.

\end{abstract}
\maketitle

\section{ Introduction}

Non-locality is one of the defining features of quantum mechanics
qualitatively differentiating it from classical physics \cite{BrunnerCavalcantiPironioEtAl-Bellnonlocality-2014}. 
Apart from
its foundational importance, scientists have recently realized that quantum
non-locality is also an extremely valuable resource enabling various tasks,
such as quantum key distribution \cite{Ac'inBrunnerGisinEtAl-Device-IndependentSecurityof-2007,
VaziraniVidick-FullyDevice-IndependentQuantum-2014} or randomness expansion and
amplification \cite{
Colbeck-QuantumAndRelativistic-2009,
PironioAc'inMassarEtAl-Randomnumberscertified-2010,
vazirani2012certifiable,
GallegoMasanesEtAl-Fullrandomnessfrom-2013,
RamanathanBrandaoHorodeckiEtAl-Randomnessamplificationagainst-2015,
BoudaPawlowskiPivoluskaEtAl-Device-independentrandomnessextraction-2014,
PleschPivoluska-Device-independentrandomnessamplification-2014,
PivoluskaPlesch-DeviceIndependentRandom-2014}. All these applications use a unifying feature of
quantum mechanics -- namely its possibility to provide the experimentalist 
results that exhibit super-classical correlations. Measurements on distant
parts of a quantum system can, if performed in a specific way, produce results
that are not reproducible by any classical system, even with the help of pre-shared
information. Since the seminal work of Bell \cite{Bell-Einstein-Podolsky-Rosenparadox-1964}, who first realized this
fact, a long line of research was devoted both to experimental realization of
different tests of quantumness (including the recent loophole-free Bell
experiment \cite{HensenBernienDr'eauEtAl-Experimentalloophole-freeviolation-2015}) and its theoretical implications.

One of the recent utilizations of quantum super-correlations is the idea of
Device Independence. As quantum devices are capable of producing a different
flavour of correlations then purely classical ones, the existence of
these kind of correlations (\textit{a.~k.~a.} violating some kind of Bell inequality)
certifies a quantum nature of the experiment performed. Thus by observing
the output data of an experiment and relating it to its input, one is in
principle able to conclude quantum nature of the devices, without any need of
knowing or testing the inner workings of the devices. And as quantum
measurements providing super-classical correlations are inevitably connected
with randomness of the outcomes, an experiment can simultaneously check
``quantumness'' of the devices and provide randomness. This approach is called
Device Independence \cite{BrunnerCavalcantiPironioEtAl-Bellnonlocality-2014}, and stands in the spotlight of recent research
in the area of quantum information.

\begin{figure}[h]
\includegraphics[scale = 2.0]{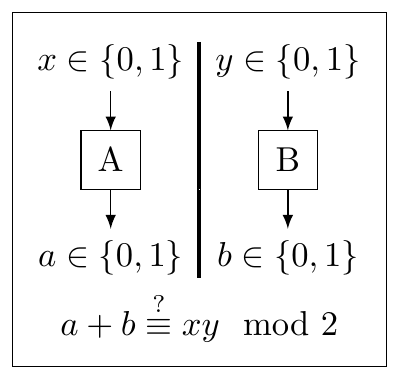}\caption{Two non-communicating
players Alice ($A$) and Bob ($B$) get one bit inputs $x$ and $y$ each, chosen at
random. Their goal it to produce two outputs $a$ and $b$ such that $a+b \equiv xy
\mod 2$.}%
\label{fig:CHSH}%
\end{figure}

Arguably the simplest and most studied generalization of the original Bell
setting is the Clauser-Horne-Shimony-Holt ($\mathrm{CHSH}$) setting \cite{ClauserHorneShimonyEtAl-ProposedExperimentto-1969}, 
where two
experimentalists choose one out of two possible binary measuremets on their
part of the system. The setting can be rephrased into a language of games,
where two non-communicating players, Alice and Bob, both receive a uniformly
chosen single bit input $x$ and $y$ respectively and their goal is to produce
single bit outputs $a$ and $b$, such that $a+b\equiv xy\mod 2$ (see Fig.
\ref{fig:CHSH}).

It is well known that classical players can win this game with probability no more than $75\%$. 
The strategy achieving this is trivial, consisting of outputting a
$0\,$by both Alice and Bob, irrespectively on the inputs. Utilizing quantum
mechanics, players can share a maximally entangled state of two qubits and perform
a suitable measurement (dependent on the input) on their respective qubit. In such a way they
 can increase the probability of wining the game up to $\frac
{2+\sqrt{2}}{4}\approx85\%$. This fact can be utilized to perform
device-independent experiments.

\begin{figure}[h]
\includegraphics[scale = 2.0]{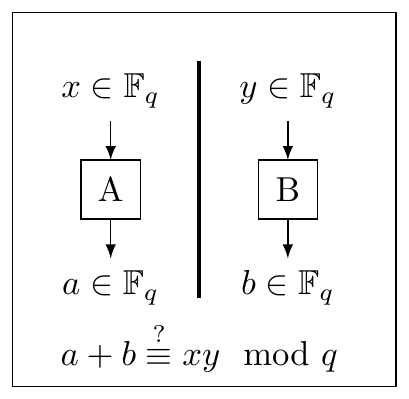}\caption{Two non-communicating
players Alice ($A$) and Bob ($B$) get inputs $x$ and $y$ chosen at random from
a finite field ${\mathbb{F}}_{q}$ with prime $q$. Their goal it to produce two outputs
$a,b\in{\mathbb{F}}_{q}$ respectively, such that $a+b\equiv xy \mod q$.}%
\label{fig:CHSHq}%
\end{figure}

With the standard $\mathrm{CHSH}$ setting, in a single round of the protocol
only two bits are produced, where only one of them can be utilized due to
the correlation with the other output bit. Therefore there appears a natural question
if and how one might produce more bits in a single experimental run. This can be easily
achieved by allowing Alice and Bob to receive an input from a higher alphabet
and also producing a more complicated result. A straightforward
generalization is a $\mathrm{CHSH}_{q}$ game, where the dimensionality of
both inputs and outputs is limited to a prime $q$ (see Fig. \ref{fig:CHSHq}). In this
case, the winning condition states $a+b\equiv xy\mod q.$ However, to be
useful for device independent experiments, the probability of winning the
game with a quantum strategy must be higher than the probability with purely
classical systems. Therefore, bounds for these probabilities are of utmost
importance for its possible use. In this paper we provide a constructive lower
bound for the probability of winning a $\mathrm{CHSH}_{q}$ game using purely
classical systems.

The paper is organized as follows. In the second section we formally define
the $\mathrm{CHSH}_{q}$ game and review the existing bounds for both classical
and quantum strategies. In the third section we relate the problem of finding
classical strategies to $\mathrm{CHSH}_q$ games to solving the problem of
point-line incidences. In the section four we introduce our classical strategy
and prove its efficiency, whereas in the last section we conclude by
discussing the results obtained.

\section{General $\mathrm{CHSH}_{q}$ games}

Formally, with a non-local game $G$, we associate two values: a classical
probability of winning $\omega(G)$ and a quantum probability of winning
$\omega^{\ast}(G)$. The non-local properties of quantum theory are
demonstrated by the fact that $\omega^{\ast}(G)>\omega(G)$. In case of the
standard binary $\mathrm{CHSH}$ game we have $\omega(\mathrm{CHSH})=0.75$ and
$\omega^{\ast}(\mathrm{CHSH})\frac{2+\sqrt{2}}{2}\approx85\%$. Both these
values are known exactly and for both quantum and classical case there exist a constructive
strategy that achieves this bound and is efficient to calculate. In fact, 
the classical strategy is fully trivial with a constant output, whereas
the quantum strategy consists of selecting a proper measurement setting
given by the binary input and providing the measurement result as the output.

The binary $\mathrm{CHSH}$ game can be generalized in the following way. Both
Alice and Bob receive inputs $x,y\in\mathbb{F}_{q}$, \textit{i.e.} a finite
field which exist for any $q$ being a prime power.   
Their goal is to produce outputs
$a,b\in\mathbb{F}_{q}$, such that $a+b\equiv xy,$ where both sum and product
are operations of the corresponding field. We will denote a game with inputs
in $\mathbb{F}_{q}$ as $\mathrm{CHSH}_{q}$. 

Note that for this section and the next section we consider the most general case of $q$ being a prime power,
however in section \ref{sec:Strategy}, we switch to prime $q$'s only. The reason for this is the fact that
in prime finite fields both addition and multiplication are very intuitive -- they are just addition and multiplication
modulo $q$, which is vital for the proofs.

\subsection{Quantum bound}

Contrary to the binary $\mathrm{CHSH}$ game, neither the exact value
$\omega^{\ast}({\mathrm{CHSH}}_{q})$, nor a strategy obtaining the optimal value is
known. The only existing result due to \cite{BavarianShor-InformationCausalitySzemeredi-Trotter-2015} introduces an upper bound for
the quantum probability
\[
\omega^{\ast}({\mathrm{CHSH}}_{q})\leq\frac{1}{q}+\frac{q-1}{q}\frac{1}%
{\sqrt{q}}=\frac{1}{\sqrt{q}}+\frac{1}{q}-\frac{1}{q\sqrt{q}}.
\]
This fact has two important consequences. The first is that being it an upper
bound, we will not be able to show $\omega^{\ast}({\mathrm{CHSH}}_{q}%
)>\omega({\mathrm{CHSH}}_{q})$ and thus the usefulness of use of ${\mathrm{CHSH}}_{q}$ for device
independent experiments. The second consequence is that even the upper bound
decreases with $\frac{1}{\sqrt{q}}$ in the leading order with large $q$. Thus,
even if the tightness of this bound and a classical-quantum gap could be shown
in the future, the statistics of successful outcomes would decreasing with $q$ and many 
experimental runs would be needed.

\subsection{Classical bounds}

With classical bounds the situation is slightly better. There exists an upper
bound in the form
\[
\omega({\mathrm{CHSH}}_{q})=%
\begin{array}
[c]{ll}%
O\left(  q^{-\frac{1}{2}-\varepsilon}\right)   & \quad\text{for $q=p^{2k+1}$},
\end{array}
\]
where $p$ is a prime, $k\geq1$ and $\varepsilon>0$ is a constant. It is
only valid for the case of an odd prime power, but still could serve for a proof
of a classical \ -- quantum gap if the quantum bound would be proven tight.

There also exists a set of lower bounds in the form
\[
\omega({\mathrm{CHSH}}_{q})=\left\{
\begin{array}
[c]{ll}%
\Omega\left(  q^{-\frac{1}{2}}\right)   & \quad\text{for $q=p^{2k}$}\\
\Omega\left(  q^{-\frac{2}{3}}\right)   & \quad\text{for $q=p^{2k+1}$}%
\end{array}
\right.  .
\]
We see that for $q$ being an even power prime the lower bound is higher than
for odd powers and thus for all values of $q$ there is a significant gap
between the lower or upper (partly non-existent) bounds.

Even more importantly and perhaps surprisingly, these lower bounds are not
connected with any concrete strategy. Quantum strategies existing so far are
limited to different heuristics (\textit{e.g.} trying to maximize the winning
probability over all measurements of the maximally entangled bipartite state),
random searches and numerics \cite{JiLeeLimEtAl-MultisettingBellinequality-2008,LiangLimDeng-Reexaminationofmultisetting-2009}. 
Best known classical strategies so far
obtained only $\omega({\mathrm{CHSH}}_{q})=\Omega\left(  \frac{1}{q}\right) $
\cite{LiangLimDeng-Reexaminationofmultisetting-2009}, which corresponds to a trivial strategy (both Alice and Bob output
$0$ irrespective on their input and win if either $x=0$ or $y=0$, thus in
$2q-1$ out of $q^{2}$ cases).

In this paper we present the first constructive classical strategy for the
${\mathrm{CHSH}}_{q}$ game with the probability of winning
$\Omega\left(  q^{-\frac{2}{3}}\right)  $ for $q$ being a prime. With this
strategy we close the gap between constructive strategies and existence
bounds. To be able to present details of the proof, we first relate the
problem of classical ${\mathrm{CHSH}}_{q}$ game strategies to a well-known
problem of point-line incidences.

\section{Point-line incidences and classical strategies for the
${\mathrm{CHSH}}_{q}$ game}

Every classical strategy of ${\mathrm{CHSH}}_{q}$ can be written as a convex
combination of deterministic strategies, which can be written down as two
functions -- $a: \mathbb{F}_{q} \mapsto\mathbb{F}_{q}$ representing the
strategy of Alice and $b: \mathbb{F}_{q} \mapsto\mathbb{F}_{q}$ representing
Bob's strategy.

The winning condition now states
\begin{equation}
a(x)+b(y) = xy,\label{Podmienka1}%
\end{equation}
which can be rewritten into a form
\[
a(x) =  xy-b(y),
\]
where all additions and multiplications are operations of the finite field $\F_q$.
Note that in this form Alice's strategy can be seen as a set of points
$P=\left\{  (x,a(x))\in\mathbb{F}_{q}^{2}\right\}  $ and Bob's strategy can be
seen as a set of lines $L=\left\{  l_{y,-b(y)}\subseteq\mathbb{F}_{q}%
^{2}\right\}  $, where a line $l_{y,-b(y)}$ contains all points $(g,h)\in
\mathbb{F}_{q}^{2}$, such that $h=yg-b(y)$. Note that the strategy of Alice
and Bob is successful for input $x,y$ if the point specified by a vector
$(x,a(x))$ lies on a the line specified by $(y,-b(y))$. Assuming uniform
choice of the input pairs, the strategy of Alice and Bob is the more
successful, the more of the points of $P$ lie on the lines in $L$. Thus one
can reformulate the problem of the best strategy for Alice and Bob to a
problem of finding $q$ points and $q$ lines with the highest number of
incidences. This is a well known and hard problem, even for general sets of
points and lines \cite{Dvir-IncidenceTheoremsand-2010}. However, in order to be able to map a set of points
and lines to a classical strategy for ${\mathrm{CHSH}}_{q}$, two more
conditions need to be fulfilled:

\begin{itemize}
\item No two points lie on the same vertical line (have the same $x$);

\item No two lines have the same slope $y$.
\end{itemize}
Violation of these conditions would make the strategy ambiguous, since it
would assign more than one possible output to some inputs $x,y$.

Let us label the number of point-line incidences by $I$. The fraction of
inputs for which Alice and Bob can produce a correct outcome is given by
$\frac{I}{q^{2}}$, which, with an assumption of uniform choice of input pairs
$(x,y)$, also gives the probability of winning the ${\mathrm{CHSH}}_{q}$ game.

\section{Strategy}\label{sec:Strategy}

In this section we construct a strategy for Alice and Bob to win the $\CHSH_q$ game
for prime $q$.
We do so by showing an explicit construction for $q$ points and $q$ lines 
with $I=\frac{1}{22}q^{4/3}$ and
thus a fraction of correct outcomes $\frac{1}{22}q^{-2/3}$. We achieve this by
selecting a specific set of points and lines not obeying the unambiguity
conditions stated before, but having a large number of mutual incidences. Then
we perform a transformation that will remove the ambiguities at the cost of
removing a portion of the lines and points we started with. In what follows we
will use the letter $p$ instead of $q$ to stress that the sums and products
are being performed in a field ${\mathbb{F}}_{p}$ of prime order $p$. We will
also use the symbol $\equiv$ in equations valid modulo $p$ (unless explicitly
a different modulo is stated) and symbol $=$ in standard integer/rational equations.

\subsection{Selection of points and lines}

We define the following quantities%

\begin{align}
p_{1} &  =2\left\lfloor \frac{p^{1/3}}{2}\right\rfloor ,\label{p1}\\
p_{2} &  =2\left\lfloor \frac{p}{2p_{1}}\right\rfloor .\label{p2}%
\end{align}
We see that both $p_{1}$ and $p_{2}$ are even and the following inequalities hold:%

\begin{align}
p_{1}^{2} &  <p_{2}\label{p_bounds_1}\\
p-2p_{1} &  <p_{1}p_{2}<p.\label{p_bounds_2}%
\end{align}
Now we define a set of $p_{1}p_{2}$ points by all points with coordinates
$(x,a)$ and

\begin{align}
x &  \in\left\langle 0,p_{1}\right)  \label{Range_body}\\
a &  \in\left\langle 0,p_{2}\right)  .\nonumber
\end{align}
We also define $\frac{p_{1}p_{2}}{4}$ lines in the form $(y,b)$ with%

\begin{align}
y &  \in\left\langle 0,\frac{p_{1}}{2}\right)  \label{Range_priamky}\\
b &  \in\left\langle 0,\frac{p_{2}}{2}\right)  .\nonumber
\end{align}
It is easy to see that each line contains exactly $p_{1}$ points with
different $x$ coordinates. The highest $a$ reached by the lines for $x<p_{1}$
is $\frac{p_{2}}{2}-1+\left(  \frac{p_{1}}{2}-1\right)  \left(  p_{1}%
-1\right)  <p_{2}-1$ and thus the number of incidences within this set is exactly
\[
I=\frac{p_{1}^{2}p_{2}}{4},
\]
which is roughly $\frac{p^{\frac{4}{3}}}{4}$.

\subsection{Transformation}

Now we perform the following transformation of both points and lines:%
\begin{align}
(x,a) &  \rightarrow\left(  \frac{1}{p_{2}x-a},1+\frac{2a}{p_{2}x-a}\right)
\label{Transformations}\\
(y,b) &  \rightarrow\left(  \frac{2p_{2}b}{p_{2}-y},\frac{p_{2}+y}{p_{2}%
-y}\right)  ,\nonumber
\end{align}
where all sums and products are performed in ${\mathbb{F}}_{p}$ and division is
understood as multiplication by the inverse element. Transformation is well
defined for all the points but $(0,0)$ and all the lines. With a bit of
technical exercise one can see that the transformed points lie on a
transformed line if and only if the original points did. It is also easy to
see that we have successfully removed all the ambiguity in points, as
$p_{2}x-a$ is different for all pairs of $(x,a)$ satisfying (\ref{Range_body})
and so is the inverse element. Therefore, we have a new set of $\left(
p_{1}p_{2}-1\right)  $ points that all have different $x$ coordinates.

The situation of lines is much different. The slope is defined by the fraction
$\frac{2p_{2}b}{p_{2}-y}$, for which it is not easy to see how many different
values it can acquire in ${\mathbb{F}}_{p}$. Here we will show that among the
$\frac{p_{1}p_{2}}{4}$ lines transformed according to (\ref{Transformations})
there will be at least $\frac{p_{1}p_{2}}{20}$ with different slopes.

\subsection{Identifying ambiguities}

In order to prove the result, we will sum up all the lines that share a
slope with another line and show that there aren't too many of them. In fact
we could leave one of the lines sharing a slope with another lines and remove
all the rest, but instead we will remove all of them. This makes the procedure
redundant, but easier to tackle and does not influence the final results by more
than a constant.

We will work with the equation $k^{\prime}\equiv\frac{2p_{2}b}{p_{2}-y}$,
which, after the substitution $k\equiv\frac{2p_{2}}{k^{\prime}}$, is
equivalent to the equation

\begin{equation}
kb\equiv p_{2}-y.\label{equation_k}%
\end{equation}
We will search for values of $k$ for which there exists more than one solution
of $y$ and $b$ within the given range (\ref{Range_priamky}). To do so, we can
visualize the situation as follows: We start from the element $0$ in the field
of length $p$ (corresponding to $b=0$ on the left-hand side of
(\ref{equation_k})) and make steps of length $k$ (corresponding to increasing
$b$). We are allowed to make up to $\frac{p_{2}}{2}-1$ steps (due to
(\ref{Range_priamky})) and are seeking for cases when we ``visit'' the interval

\begin{equation}
\left(  p_{2}-\frac{p_{1}}{2},p_{2}\right\rangle \label{Interval}%
\end{equation}
more than once, as this is the interval of values the right-hand side of
(\ref{equation_k}) can acquire. This can happen in two principally different cases:

\begin{itemize}
\item $k<\frac{p_{1}}{2}$ and thus the interval is repeatedly visited within
subsequent steps
\item $k>2p_{1}$ and the interval is visited after one or more cycles within
the field.
\end{itemize}
For $\frac{p_{1}}{2}\leq k\leq2p_{1}$, the size of the step is larger than the
interval we are trying to hit, therefore we cannot visit the interval twice without
at least one cycle in the field, yet the step is too short to finish a single cycle
within the field. Thus for $\frac{p_{1}}{2}\leq k\leq2p_{1}$ there cannot
exist more than one solution of (\ref{equation_k}).

\subsubsection{Small steps}

If $k<\frac{p_{1}}{2},$ the analysis is very simple. We can upper bound the
number of solutions for each $k$ to $\frac{p_{1}}{2}$ and thus the number of
repeated solutions to $R_{\mathrm{small}}=\frac{p_{1}^{2}}{4}$. This bound is in fact
very loose, but for large $p$ is fully satisfactory.

\subsubsection{Large steps}

The second case is more complicated. Here we know that the left hand side of
equation (\ref{equation_k}) is $0$ \ for $b=0$. Let $b_{1}$ be the smallest
$b$ such that (\ref{equation_k}) holds for a given $k$ and let $b_{2}$ be the
next $b$ for which (\ref{equation_k}) holds. Let

\begin{equation}
d\equiv k\left(  b_{2}-b_{1}\right)  .\label{d}%
\end{equation}
We now define%

\begin{equation}
\delta=\left\{
\begin{array}
[c]{ll}%
d & \quad\text{for $d<p/2$}\\
d-p & \quad\text{for $d>p/2$}.%
\end{array}
\right.  \label{delta}%
\end{equation}
Here $\delta$ is an integer, thus $\delta$ can acquire both positive and negative
values and therefore $|\delta|$ is the standard absolute value. It is easy to
see that

\begin{equation}
\left\vert \delta\right\vert <\frac{p_{1}}{2},\label{range_delta}%
\end{equation}
due to the limited width of the interval (\ref{Interval}). This condition
means that before visiting the interval $\left(  p_{2}-\frac{p1}{2}%
,p_{2}\right\rangle $ twice, we have to visit also the interval $\left(
-\frac{p1}{2},\frac{p1}{2}\right\rangle $ in point $d\equiv\delta$ once again
after staring from $0$.

\begin{figure}
\includegraphics[scale = 1.5]{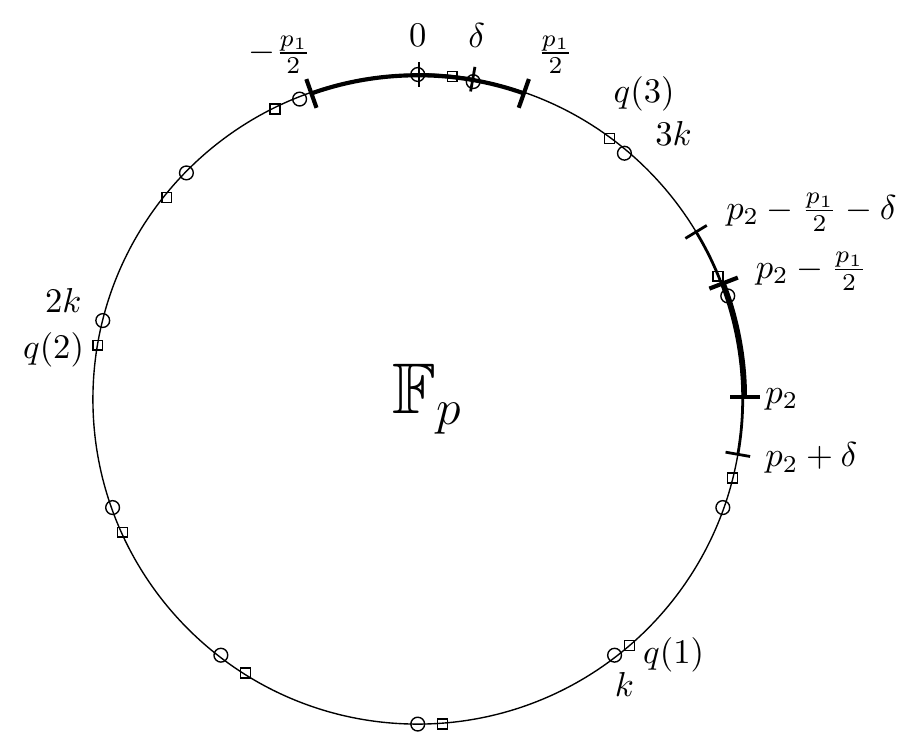}
\caption{In the Figure we show a visualization of different quantities in the field. 
Circles are used to denote the actual points in the field in the form $r.k$, whereas the squares are used for 
approximations of the points $q(r)$. }
\label{circle}
\end{figure}

Let us now define

\begin{equation}
l\equiv\frac{d}{k}.\label{l}%
\end{equation}
This means that after $l$ steps of length $k$ we visit the point
$d\equiv\delta$. Switching back to integers, this means that there exists a
positive $s$ such that

\begin{equation}
kl=sp+\delta.\label{kl}%
\end{equation}
As $k<p$ and $d<p$, clearly $0\leq s\leq l$ and as $k>2p_{1}$, $s>0$.
We can also write

\begin{equation}
k=\frac{sp+\delta}{l}.\label{def_s}%
\end{equation}
Now it is easy to see that the points in the field visited in $r^{\mathrm{th}%
}$ step $(0<r<l)$ have the form

\begin{equation}
rk=r\frac{sp+\delta}{l}.\label{def_r}%
\end{equation}
Let us now define a set of rational numbers

\begin{equation}
Q=\left\{\left.  q\frac{p+\delta}{l}\right|0<q<l\right\}\label{def_q}.%
\end{equation}
For the specific case $s=1$, elements of $Q$ are natural numbers and for each
$q$ they exactly define elements of ${\mathbb{F}}_{p}$ visited by the
$q^{\mathrm{th}}$ step of length $k$. In all the other cases we want to relate
the $r^{\mathrm{th}}$ visited element of the field with a specific element of
$Q$. We do it as follows -- the $r^{\mathrm{th}}$ visited point is associated
with element of $Q$ defined by

\begin{equation}
q(r):=rs\operatorname{mod}l\label{def_q_2}.%
\end{equation}
Therefore the element of ${\mathbb{F}}_{p}$ $r\frac{sp+\delta}{l}$ is
associated with (see also Fig. \ref{circle})

\begin{equation}
q(r)\frac{p+\delta}{l}=\left(  rs\operatorname{mod}l\right)  \frac{p+\delta
}{l}.\label{association}%
\end{equation}
It is easy to see that for all $q$ it holds $q\frac{p+\delta}{l}<p.$ What is
not trivial to see is that the relation between $q(r)$ and $r$ is a bijection:
In order to show a contradiction, consider $r_{1}<r_{2}$ for which
$q(r_{1})=q(r_{2})=q$. Then $l$ divides $s(r_{2}-r_{1})$ and the step
$r_{2}-r_{1}$ points to

\begin{equation}
(r_{2}-r_{1})k=\frac{(r_{2}-r_{1})sp+(r_{2}-r_{1})\delta}{l}\equiv\delta
\frac{r_{2}-r_{1}}{l}.\label{eq_delta1}%
\end{equation}
Note that since the left hand side of the equation is an integer, so is it's
right hand side. Additionally, since $r_{2}-r_{1}<l$, we also have $\left\vert
\delta\right\vert >\left\vert \delta\frac{r_{2}-r_{1}}{l}\right\vert $. This
would mean that before reaching the point $\delta$ in $l$ steps, we would
reach a point $\delta\frac{r_{2}-r_{1}}{l}$, which is closer to zero than
$\delta$, in less than $l$ steps, which is a contradiction with the definition
of $\delta$. Thus we can conclude that there is a one-to-one correspondence
between $r$ and $q(r)$.

Now we calculate the difference between the points defined by $r$ and by
$q(r)$:

\begin{equation}
r\frac{sp+\delta}{l}-q(r)\frac{p+\delta}{l}=p\frac{rs-q(r)}{l}+\delta
\frac{r-q(r)}{l}\equiv\delta\frac{r-q(r)}{l},\label{eq_delta2}%
\end{equation}
where in the second equivalence we used (\ref{def_q_2}). As both $r$ and
$q(r)$ are non-negative and smaller than $l$, the absolute value of this
distance is smaller than $|\delta|$.

This has an important consequence. We can conclude that the points visited by
walking through the field with $l$ steps of length $k$ (defined in
(\ref{def_r})) are elements of the field -- natural numbers that do not differ
by more than $|\delta|$ from the rational numbers defined in (\ref{def_q}) for
$0<q<l$. So the interval (\ref{Interval}) was possibly visited only if the
newly defined points $Q$ fit into a larger interval

\begin{equation}
q\frac{p+\delta}{l}\in\left(  p_{2}-\frac{p_{1}}{2}-|\delta|,p_{2}%
+|\delta|\right\rangle \label{interval_delta}%
\end{equation}
for some value of $0<q<l$. The last condition can be rewritten in a form of an
inequality%

\begin{equation}
p_{2}-\frac{p_{1}}{2}-|\delta|<q\frac{p+\delta}{l}\leq p_{2}+|\delta
|\text{.}\label{inequality}%
\end{equation}

Now the task is to find for which values of $l$ there exist a suitable $q$
that fulfills this inequality. For each of these values of $l$ we will have to
remove the number of possible repeated solutions, which is defined as a
minimum of $\frac{p_{2}}{2l}$ due to the limited number of steps allowed (as
each return to the interval costs $l$ steps and only $\frac{p_{2}}{2}$ steps
are available) and $\frac{p_{1}}{2|\delta|}$ due to the narrowness of the
interval (if the first return to the interval was $\delta$ away from the first
visit, the second return will be $2\delta$ away etc. and the interval is only
$\frac{p_{1}}{2}$ broad).

We can rephrase the task also in a slightly different way: for each natural $q<l$,
we will find the number of different values of $l$ that fulfill
(\ref{inequality}) and for each of them calculate the number of repeated
solutions. One very important observation is that (\ref{inequality}) can only
have solutions for $q\leq\frac{p_{1}}{2}$. This is easy to see from the fact
that the middle part of (\ref{inequality}) needs to be smaller than or equal
to $p_{2}+|\delta|$, and as it is always smaller than $p$, we can limit
ourselves to solutions within natural numbers, without taking into account
field properties. As $l$ is limited to $\frac{p_{2}}{2}$, using
(\ref{p_bounds_1}) we get $q\leq\frac{p_{1}}{2}$.

Let us define $l_{q}$ as the largest $l$ for which (\ref{inequality}) is
satisfied. Then it holds:%

\begin{equation}
l_{q}<q\frac{p+\delta}{p_{2}-\frac{p_{1}}{2}-|\delta|}.\label{lq}%
\end{equation}
Let us also define $l_{q}-x_{q}$ as the smallest $l$ for which
(\ref{inequality}) is satisfied. For $l_{q}-x_{q}$ to solve the inequality, it
must hold

\begin{equation}
l_{q}\left(  p_{2}-\frac{p_{1}}{2}-|\delta|\right)  <(l_{q}-x_{q}%
)(p_{2}+|\delta|)\label{xq_definition}%
\end{equation}
and thus
\begin{equation}
x_{q}<l_{q}\frac{\frac{p_{1}}{2}+2|\delta|}{p_{2}+|\delta|}.\label{xq_bound}%
\end{equation}
We identify $x_{q}$ as the number of different $l$s that can solve
(\ref{inequality}) for a fixed value of $q$.

\subsection{Removing ambiguities}

Now we can calculate the upper bound of repeated solutions for a specific
$\delta$%
\begin{equation}
R_{\delta}=\sum_{q=1}^{p_{1}/2}x_{q}.\min\left[  \frac{p_{1}}{2|\delta|}%
,\frac{p_{2}}{2l_{q}}\right]  <\sum_{q=1}^{p_{1}/2}l_{q}\frac{\frac{p_{1}}%
{2}+|\delta|}{p_{2}+|\delta|}\min\left[  \frac{p_{1}}{2|\delta|},\frac{p_{2}%
}{2l_{q}}\right]  .\label{Rdelta1}%
\end{equation}
We get rid of the minimum in the sum by a simple trick -- as $l_{q}$ grows
with $q$, we will take the first value $\frac{p_{1}}{2|\delta|}$ for small
values of $q$ and the second value $\frac{p_{2}}{2l_{q}}$ for larger values of $q$. We choose
the breaking point to be $|\delta|$, which is roughly where the transition
takes place. Importantly, we do not need to make this decision precise, as a
wrong breaking point will only increase the value of the sum and we are
interested in an upper bound. The sum now reads 

\begin{equation}
R_{\delta}<\frac{\frac{p_{1}}{2}+|\delta|}{p_{2}+|\delta|}\left[  \sum
_{q=1}^{|\delta|}l_{q}\frac{p_{1}}{2|\delta|}+\sum_{q=|\delta|+1}^{p_{1}%
/2}\frac{p_{2}}{2}\right]  .\label{Rdelta2}%
\end{equation}
After substituting for $l_{q}$, the sums can be solved and yield

\begin{equation}
R_{\delta}<\frac{\frac{p_{1}}{2}+|\delta|}{p_{2}+|\delta|}\left[
\frac{\left(  |\delta|-1\right)  }{2}\frac{p+\delta}{p_{2}-\frac{p1}%
{2}-|\delta|}\frac{p_{1}}{2}+\left(  \frac{p_{1}}{2}-|\delta|\right)
\frac{p_{2}}{2}\right]  .\label{Rdelta3}%
\end{equation}
Using (\ref{p_bounds_1}) and (\ref{p_bounds_2}) it is easy to see that
$\frac{p+\delta}{p_{2}-\frac{p1}{2}-|\delta|}p_{1}<p_{2}$, thus

\begin{equation}
R_{\delta}<\frac{\frac{p_{1}}{2}+|\delta|}{4}\left(  p_{1}-|\delta|\right)
.\label{Rdelta4}%
\end{equation}
Now we are ready to sum all $R_{\delta}$ with $\delta$ in the interval given
by (\ref{range_delta}). As only absolute value of $\delta$ enters into the
formula and $\delta=0$ is not a valid case, we can write:%
\begin{equation}
R_{\mathrm{large}}<2\sum_{\delta=1}^{p_{1}/2}\frac{\frac{p_{1}}{2}+\delta}{4}\left(
p_{1}-\delta\right)  <p_{1}^{3}\left(  \frac{1}{8}+\frac{1}{16}-\frac{1}%
{48}\right)  =\frac{p_{1}^{3}}{6}\label{R2}.%
\end{equation}
The total number of repeated solutions is then upper bounded by
\begin{equation}
R=R_{\mathrm{small}}+R_{\mathrm{large}}=p_{1}^{2}\left( \frac{1}{2}+ \frac{p_{1}}{6}\right)
.\label{R}%
\end{equation}
As $\frac{p_{1}}{6}+\frac{1}{2}<\frac{p_{1}}{5}$ for $p_{1}>30$ (thus for
fields larger than $27000$) and $p_{2}>p_{1}^{2}$ due to (\ref{p_bounds_1}),
we can upper bound
\begin{equation}
R<\frac{p_{1}p_{2}}{5}.\label{R_bound}%
\end{equation}
So even if we remove all repeated solutions, we will stay with at least
$\frac{p_{1}p_{2}}{4}-\frac{p_{1}p_{2}}{5}=\frac{p_{1}p_{2}}{20}$ lines, each
reaching at least $(p_{1}-1)$ points (as we lost one point during the
transformation). This will lead us to
\begin{equation}
I=\frac{p_{1}p_{2}}{20}(p_{1}-1)\label{I}%
\end{equation}
point-line incidences.

Using (\ref{p_bounds_2}) we can write
\begin{equation}
I>\frac{p-2p_{1}}{20}(p_{1}-1)>\frac{pp_{1}}{20}-\frac{p_{1}^{2}}{10}-\frac
{p}{20}>\frac{pp_{1}}{21}\label{I_bound1}%
\end{equation}
for $p_{1}>30$, as $pp_{1}>42p_{1}^{2}+21p$. Further we can show that
\begin{equation}
I>\frac{p^{4/3}}{22},\label{I_bound2}%
\end{equation}
as $p_{1}>\frac{21}{22}p^{1/3}$ for $p_{1}>30$.

\subsection{Formulating the strategy}

In the previous subsection we have shown that the number of incidences is
lower bounded by $\frac{p^{4/3}}{22}$. Now we are ready to formulate the
strategy for both Alice and Bob, which will utilize this fact and lead to a
victory in the ${\mathrm{CHSH}}_{q}$ game in more then $\frac{p^{4/3}}{22}$
cases out of $p^{2}$.

For Alice, the situation is rather simple. After getting the input $x\neq0$,
she will compute the inverse element $x^{-1}$. Then she will find the solution
of an equation $x^{-1}=p_{2}x^{\prime}-a^{\prime}$ within the range
(\ref{Range_body}). To do that, she will need to calculate the inverse element
of $x$ (which can be done in an efficient way) and find the quotient ($x^{\prime}-1$)
and remainder ($p_{2}-a^{\prime}$) after dividing by $p_{2}$. If the resulting
$x^{\prime}$ and $a^{\prime}$ fit into the range (\ref{Range_body}) (this will
happen in $p_{2}p_{1}$ cases), she will compute the outcome as
$a=1+\frac{2a^{\prime}}{p_{2}x^{\prime}-a^{\prime}}$, which again involves an
efficient computation of an inverse element. In all remaining cases Alice will
return $0$, mimicking the trivial strategy.

Bob is in a slightly more complicated situation. For a given input $a$ he will
have to find the solution of $a=\frac{2p_{2}b^{\prime}}{p_{2}-y^{\prime}}$
within the range (\ref{Range_priamky}). In the worst case he will have to try
$p_{1}/2$ different values for $y^{\prime}$ and check whether $\frac{a}%
{2}-\frac{y^{\prime}}{p_{2}}<\frac{p_{2}}{2}$. If he finds a solution for $y^{\prime}$, he will
output $b=\frac{p_{2}+y^{\prime}}{p_{2}-y^{\prime}}$. In this way he will also
utilize some of the ambiguous solutions (he will keep the first $y^{\prime}$
that satisfies conditions, even if other values might as well) -- this will
potentially lead to winning the game (if the choice by Alice reflects correctly
the solution chosen by Bob), but this chance is not incorporated in the bound.
If Bob does not find a solution after trying all possible $y^{\prime}$, he
will output $0$.

Bob will need to calculate the inverse element of $p_{2}$ and $2$, which are
one-off efforts. Then he will need to perform simple inequality check up to
$p_{1}/2$ times and if successful, he will need to calculate one more inverse
element. If this would be considered still inefficient, he can adopt the
techniques from the previous subsections to find approximate values of
$\frac{y^{\prime}}{p_{2}}$ for the set of $0\leq y^{\prime}%
<\frac{p_{2}}{2}$ in advance and then test only a minor subset of $y^{\prime}$s.

\section{Conclusion}

In this paper we have provided an explicit constructive strategy for winning a
generalized ${\mathrm{CHSH}}_{q}$ game. The winning probability is lower
bounded by $\frac{p^{-2/3}}{22}$, what perfectly mimics the non-constructive
existence bound known so far.

This result is useful for potential design of device independent algorithms
based on higher alphabet ${\mathrm{CHSH}}$ games in different aspects. First,
it closes the gap between existing explicit strategies and proven existence bounds, which
helps the understanding of the nature of the problem. Second, and most
importantly, the presented result provides the first non-trivial classical
strategy for a ${\mathrm{CHSH}}$ game, where Alice and Bob need to act in a
way that depends on their input and their output is a result of a non-trivial
calculation.

There is also a set of open questions that remain. The obvious one is, how one
could generalize result presented in this paper for prime power fields. 
This is not easy, as the nature of the proof relays on the relation between addition 
and multiplication, which is unique for prime fields. Also the fact that known existence bounds
crucially depend on whether they are deployed on even or odd power prime field
suggests that any possible generalization will not be straightforward. 

More ambitious goals include the aim of finding tight bounds on classical strategies. This might, in 
accordance with suitable heuristic results for quantum strategies, lead to the possibility of 
direct use of higher-order ${\mathrm{CHSH}}_{q}$ games in experiments. The ultimate goal, naturally, 
remains to directly prove a gap between classical and quantum strategies.
\section*{Acknowledgments}
This research was supported by the Czech Science
Foundation GA\v{C}R project P202/12/1142, EU project RAQUEL, as well as project VEGA 2/0043/15.

\bibliography{Incidences}

\begin{thebibliography}{18}%
\makeatletter
\providecommand \@ifxundefined [1]{%
 \@ifx{#1\undefined}
}%
\providecommand \@ifnum [1]{%
 \ifnum #1\expandafter \@firstoftwo
 \else \expandafter \@secondoftwo
 \fi
}%
\providecommand \@ifx [1]{%
 \ifx #1\expandafter \@firstoftwo
 \else \expandafter \@secondoftwo
 \fi
}%
\providecommand \natexlab [1]{#1}%
\providecommand \enquote  [1]{``#1''}%
\providecommand \bibnamefont  [1]{#1}%
\providecommand \bibfnamefont [1]{#1}%
\providecommand \citenamefont [1]{#1}%
\providecommand \href@noop [0]{\@secondoftwo}%
\providecommand \href [0]{\begingroup \@sanitize@url \@href}%
\providecommand \@href[1]{\@@startlink{#1}\@@href}%
\providecommand \@@href[1]{\endgroup#1\@@endlink}%
\providecommand \@sanitize@url [0]{\catcode `\\12\catcode `\$12\catcode
  `\&12\catcode `\#12\catcode `\^12\catcode `\_12\catcode `\%12\relax}%
\providecommand \@@startlink[1]{}%
\providecommand \@@endlink[0]{}%
\providecommand \url  [0]{\begingroup\@sanitize@url \@url }%
\providecommand \@url [1]{\endgroup\@href {#1}{\urlprefix }}%
\providecommand \urlprefix  [0]{URL }%
\providecommand \Eprint [0]{\href }%
\providecommand \doibase [0]{http://dx.doi.org/}%
\providecommand \selectlanguage [0]{\@gobble}%
\providecommand \bibinfo  [0]{\@secondoftwo}%
\providecommand \bibfield  [0]{\@secondoftwo}%
\providecommand \translation [1]{[#1]}%
\providecommand \BibitemOpen [0]{}%
\providecommand \bibitemStop [0]{}%
\providecommand \bibitemNoStop [0]{.\EOS\space}%
\providecommand \EOS [0]{\spacefactor3000\relax}%
\providecommand \BibitemShut  [1]{\csname bibitem#1\endcsname}%
\let\auto@bib@innerbib\@empty
\bibitem [{\citenamefont {Brunner}\ \emph {et~al.}(2014)\citenamefont
  {Brunner}, \citenamefont {Cavalcanti}, \citenamefont {Pironio}, \citenamefont
  {Scarani},\ and\ \citenamefont
  {Wehner}}]{BrunnerCavalcantiPironioEtAl-Bellnonlocality-2014}%
  \BibitemOpen
  \bibfield  {author} {\bibinfo {author} {\bibfnamefont {N.}~\bibnamefont
  {Brunner}}, \bibinfo {author} {\bibfnamefont {D.}~\bibnamefont {Cavalcanti}},
  \bibinfo {author} {\bibfnamefont {S.}~\bibnamefont {Pironio}}, \bibinfo
  {author} {\bibfnamefont {V.}~\bibnamefont {Scarani}}, \ and\ \bibinfo
  {author} {\bibfnamefont {S.}~\bibnamefont {Wehner}},\ }\href {\doibase
  10.1103/RevModPhys.86.419} {\bibfield  {journal} {\bibinfo  {journal} {Rev.
  Mod. Phys.}\ }\textbf {\bibinfo {volume} {86}},\ \bibinfo {pages} {419}
  (\bibinfo {year} {2014})}\BibitemShut {NoStop}%
\bibitem [{\citenamefont {Ac\'in}\ \emph {et~al.}(2007)\citenamefont {Ac\'in},
  \citenamefont {Brunner}, \citenamefont {Gisin}, \citenamefont {Massar},
  \citenamefont {Pironio},\ and\ \citenamefont
  {Scarani}}]{Ac'inBrunnerGisinEtAl-Device-IndependentSecurityof-2007}%
  \BibitemOpen
  \bibfield  {author} {\bibinfo {author} {\bibfnamefont {A.}~\bibnamefont
  {Ac\'in}}, \bibinfo {author} {\bibfnamefont {N.}~\bibnamefont {Brunner}},
  \bibinfo {author} {\bibfnamefont {N.}~\bibnamefont {Gisin}}, \bibinfo
  {author} {\bibfnamefont {S.}~\bibnamefont {Massar}}, \bibinfo {author}
  {\bibfnamefont {S.}~\bibnamefont {Pironio}}, \ and\ \bibinfo {author}
  {\bibfnamefont {V.}~\bibnamefont {Scarani}},\ }\href {\doibase
  10.1103/PhysRevLett.98.230501} {\bibfield  {journal} {\bibinfo  {journal}
  {Phys. Rev. Lett.}\ }\textbf {\bibinfo {volume} {98}},\ \bibinfo {pages}
  {230501} (\bibinfo {year} {2007})}\BibitemShut {NoStop}%
\bibitem [{\citenamefont {Vazirani}\ and\ \citenamefont
  {Vidick}(2014)}]{VaziraniVidick-FullyDevice-IndependentQuantum-2014}%
  \BibitemOpen
  \bibfield  {author} {\bibinfo {author} {\bibfnamefont {U.}~\bibnamefont
  {Vazirani}}\ and\ \bibinfo {author} {\bibfnamefont {T.}~\bibnamefont
  {Vidick}},\ }\href {\doibase 10.1103/PhysRevLett.113.140501} {\bibfield
  {journal} {\bibinfo  {journal} {Phys. Rev. Lett.}\ }\textbf {\bibinfo
  {volume} {113}},\ \bibinfo {pages} {140501} (\bibinfo {year}
  {2014})}\BibitemShut {NoStop}%
\bibitem [{\citenamefont
  {Colbeck}(2009)}]{Colbeck-QuantumAndRelativistic-2009}%
  \BibitemOpen
  \bibfield  {author} {\bibinfo {author} {\bibfnamefont {R.}~\bibnamefont
  {Colbeck}},\ }\emph {\bibinfo {title} {Quantum And Relativistic Protocols For
  Secure Multi-Party Computation}},\ \href@noop {} {Ph.D. thesis},\ \bibinfo
  {school} {quant-ph/0911.3814} (\bibinfo {year} {2009})\BibitemShut {NoStop}%
\bibitem [{\citenamefont {{Pironio}}\ \emph {et~al.}(2010)\citenamefont
  {{Pironio}}, \citenamefont {{Ac{\'{\i}}n}}, \citenamefont {{Massar}},
  \citenamefont {{de La Giroday}}, \citenamefont {{Matsukevich}}, \citenamefont
  {{Maunz}}, \citenamefont {{Olmschenk}}, \citenamefont {{Hayes}},
  \citenamefont {{Luo}}, \citenamefont {{Manning}},\ and\ \citenamefont
  {{Monroe}}}]{PironioAc'inMassarEtAl-Randomnumberscertified-2010}%
  \BibitemOpen
  \bibfield  {author} {\bibinfo {author} {\bibfnamefont {S.}~\bibnamefont
  {{Pironio}}}, \bibinfo {author} {\bibfnamefont {A.}~\bibnamefont
  {{Ac{\'{\i}}n}}}, \bibinfo {author} {\bibfnamefont {S.}~\bibnamefont
  {{Massar}}}, \bibinfo {author} {\bibfnamefont {A.~B.}\ \bibnamefont {{de La
  Giroday}}}, \bibinfo {author} {\bibfnamefont {D.~N.}\ \bibnamefont
  {{Matsukevich}}}, \bibinfo {author} {\bibfnamefont {P.}~\bibnamefont
  {{Maunz}}}, \bibinfo {author} {\bibfnamefont {S.}~\bibnamefont
  {{Olmschenk}}}, \bibinfo {author} {\bibfnamefont {D.}~\bibnamefont
  {{Hayes}}}, \bibinfo {author} {\bibfnamefont {L.}~\bibnamefont {{Luo}}},
  \bibinfo {author} {\bibfnamefont {T.~A.}\ \bibnamefont {{Manning}}}, \ and\
  \bibinfo {author} {\bibfnamefont {C.}~\bibnamefont {{Monroe}}},\ }\href
  {\doibase 10.1038/nature09008} {\bibfield  {journal} {\bibinfo  {journal}
  {Nature}\ }\textbf {\bibinfo {volume} {464}},\ \bibinfo {pages} {1021}
  (\bibinfo {year} {2010})}\BibitemShut {NoStop}%
\bibitem [{\citenamefont {Vazirani}\ and\ \citenamefont
  {Vidick}(2012)}]{vazirani2012certifiable}%
  \BibitemOpen
  \bibfield  {author} {\bibinfo {author} {\bibfnamefont {U.}~\bibnamefont
  {Vazirani}}\ and\ \bibinfo {author} {\bibfnamefont {T.}~\bibnamefont
  {Vidick}},\ }in\ \href@noop {} {\emph {\bibinfo {booktitle} {Proceedings of
  the 44th symposium on Theory of Computing}}}\ (\bibinfo {organization}
  {ACM},\ \bibinfo {year} {2012})\ pp.\ \bibinfo {pages} {61--76}\BibitemShut
  {NoStop}%
\bibitem [{\citenamefont {{Gallego}}\ \emph {et~al.}(2013)\citenamefont
  {{Gallego}}, \citenamefont {{Masanes}}, \citenamefont {{de la Torre}},
  \citenamefont {{Dhara}}, \citenamefont {{Aolita}},\ and\ \citenamefont
  {{Ac{\'{\i}}n}}}]{GallegoMasanesEtAl-Fullrandomnessfrom-2013}%
  \BibitemOpen
  \bibfield  {author} {\bibinfo {author} {\bibfnamefont {R.}~\bibnamefont
  {{Gallego}}}, \bibinfo {author} {\bibfnamefont {L.}~\bibnamefont
  {{Masanes}}}, \bibinfo {author} {\bibfnamefont {G.}~\bibnamefont {{de la
  Torre}}}, \bibinfo {author} {\bibfnamefont {C.}~\bibnamefont {{Dhara}}},
  \bibinfo {author} {\bibfnamefont {L.}~\bibnamefont {{Aolita}}}, \ and\
  \bibinfo {author} {\bibfnamefont {A.}~\bibnamefont {{Ac{\'{\i}}n}}},\ }\href
  {\doibase 10.1038/ncomms3654} {\bibfield  {journal} {\bibinfo  {journal}
  {Nature Communications}\ }\textbf {\bibinfo {volume} {4}},\ \bibinfo {eid}
  {2654} (\bibinfo {year} {2013}),\ 10.1038/ncomms3654}\BibitemShut {NoStop}%
\bibitem [{\citenamefont {{Ramanathan}}\ \emph {et~al.}(2015)\citenamefont
  {{Ramanathan}}, \citenamefont {{Brand{\~a}o}}, \citenamefont {{Horodecki}},
  \citenamefont {{Horodecki}}, \citenamefont {{Horodecki}},\ and\ \citenamefont
  {{Wojew{\'o}dka}}}]{RamanathanBrandaoHorodeckiEtAl-Randomnessamplificationagainst-2015}%
  \BibitemOpen
  \bibfield  {author} {\bibinfo {author} {\bibfnamefont {R.}~\bibnamefont
  {{Ramanathan}}}, \bibinfo {author} {\bibfnamefont {F.~G.~S.~L.}\ \bibnamefont
  {{Brand{\~a}o}}}, \bibinfo {author} {\bibfnamefont {K.}~\bibnamefont
  {{Horodecki}}}, \bibinfo {author} {\bibfnamefont {M.}~\bibnamefont
  {{Horodecki}}}, \bibinfo {author} {\bibfnamefont {P.}~\bibnamefont
  {{Horodecki}}}, \ and\ \bibinfo {author} {\bibfnamefont {H.}~\bibnamefont
  {{Wojew{\'o}dka}}},\ }\href@noop {} {\bibfield  {journal} {\bibinfo
  {journal} {ArXiv e-prints}\ } (\bibinfo {year} {2015})},\ \Eprint
  {http://arxiv.org/abs/1504.06313} {arXiv:1504.06313 [quant-ph]} \BibitemShut
  {NoStop}%
\bibitem [{\citenamefont {Bouda}\ \emph {et~al.}(2014)\citenamefont {Bouda},
  \citenamefont {Paw\l{}owski}, \citenamefont {Pivoluska},\ and\ \citenamefont
  {Plesch}}]{BoudaPawlowskiPivoluskaEtAl-Device-independentrandomnessextraction-2014}%
  \BibitemOpen
  \bibfield  {author} {\bibinfo {author} {\bibfnamefont {J.}~\bibnamefont
  {Bouda}}, \bibinfo {author} {\bibfnamefont {M.}~\bibnamefont {Paw\l{}owski}},
  \bibinfo {author} {\bibfnamefont {M.}~\bibnamefont {Pivoluska}}, \ and\
  \bibinfo {author} {\bibfnamefont {M.}~\bibnamefont {Plesch}},\ }\href
  {\doibase 10.1103/PhysRevA.90.032313} {\bibfield  {journal} {\bibinfo
  {journal} {Phys. Rev. A}\ }\textbf {\bibinfo {volume} {90}},\ \bibinfo
  {pages} {032313} (\bibinfo {year} {2014})}\BibitemShut {NoStop}%
\bibitem [{\citenamefont {Plesch}\ and\ \citenamefont
  {Pivoluska}(2014)}]{PleschPivoluska-Device-independentrandomnessamplification-2014}%
  \BibitemOpen
  \bibfield  {author} {\bibinfo {author} {\bibfnamefont {M.}~\bibnamefont
  {Plesch}}\ and\ \bibinfo {author} {\bibfnamefont {M.}~\bibnamefont
  {Pivoluska}},\ }\href {\doibase
  http://dx.doi.org/10.1016/j.physleta.2014.08.007} {\bibfield  {journal}
  {\bibinfo  {journal} {Physics Letters A}\ }\textbf {\bibinfo {volume}
  {378}},\ \bibinfo {pages} {2938 } (\bibinfo {year} {2014})}\BibitemShut
  {NoStop}%
\bibitem [{\citenamefont {Pivoluska}\ and\ \citenamefont
  {Plesch}(2014)}]{PivoluskaPlesch-DeviceIndependentRandom-2014}%
  \BibitemOpen
  \bibfield  {author} {\bibinfo {author} {\bibfnamefont {M.}~\bibnamefont
  {Pivoluska}}\ and\ \bibinfo {author} {\bibfnamefont {M.}~\bibnamefont
  {Plesch}},\ }\href {\doibase 10.2478/apsrt-2014-0006} {\bibfield  {journal}
  {\bibinfo  {journal} {Acta Physica Slovaca}\ }\textbf {\bibinfo {volume}
  {64}},\ \bibinfo {pages} {600 } (\bibinfo {year} {2014})}\BibitemShut
  {NoStop}%
\bibitem [{\citenamefont
  {Bell}(1964)}]{Bell-Einstein-Podolsky-Rosenparadox-1964}%
  \BibitemOpen
  \bibfield  {author} {\bibinfo {author} {\bibfnamefont {J.~S.}\ \bibnamefont
  {Bell}},\ }\href@noop {} {\bibfield  {journal} {\bibinfo  {journal}
  {Physics}\ }\textbf {\bibinfo {volume} {1}},\ \bibinfo {pages} {195}
  (\bibinfo {year} {1964})}\BibitemShut {NoStop}%
\bibitem [{\citenamefont {{Hensen}}\ \emph {et~al.}(2015)\citenamefont
  {{Hensen}}, \citenamefont {{Bernien}}, \citenamefont {{Dr{\'e}au}},
  \citenamefont {{Reiserer}}, \citenamefont {{Kalb}}, \citenamefont {{Blok}},
  \citenamefont {{Ruitenberg}}, \citenamefont {{Vermeulen}}, \citenamefont
  {{Schouten}}, \citenamefont {{Abell{\'a}n}}, \citenamefont {{Amaya}},
  \citenamefont {{Pruneri}}, \citenamefont {{Mitchell}}, \citenamefont
  {{Markham}}, \citenamefont {{Twitchen}}, \citenamefont {{Elkouss}},
  \citenamefont {{Wehner}}, \citenamefont {{Taminiau}},\ and\ \citenamefont
  {{Hanson}}}]{HensenBernienDr'eauEtAl-Experimentalloophole-freeviolation-2015}%
  \BibitemOpen
  \bibfield  {author} {\bibinfo {author} {\bibfnamefont {B.}~\bibnamefont
  {{Hensen}}}, \bibinfo {author} {\bibfnamefont {H.}~\bibnamefont {{Bernien}}},
  \bibinfo {author} {\bibfnamefont {A.~E.}\ \bibnamefont {{Dr{\'e}au}}},
  \bibinfo {author} {\bibfnamefont {A.}~\bibnamefont {{Reiserer}}}, \bibinfo
  {author} {\bibfnamefont {N.}~\bibnamefont {{Kalb}}}, \bibinfo {author}
  {\bibfnamefont {M.~S.}\ \bibnamefont {{Blok}}}, \bibinfo {author}
  {\bibfnamefont {J.}~\bibnamefont {{Ruitenberg}}}, \bibinfo {author}
  {\bibfnamefont {R.~F.~L.}\ \bibnamefont {{Vermeulen}}}, \bibinfo {author}
  {\bibfnamefont {R.~N.}\ \bibnamefont {{Schouten}}}, \bibinfo {author}
  {\bibfnamefont {C.}~\bibnamefont {{Abell{\'a}n}}}, \bibinfo {author}
  {\bibfnamefont {W.}~\bibnamefont {{Amaya}}}, \bibinfo {author} {\bibfnamefont
  {V.}~\bibnamefont {{Pruneri}}}, \bibinfo {author} {\bibfnamefont {M.~W.}\
  \bibnamefont {{Mitchell}}}, \bibinfo {author} {\bibfnamefont
  {M.}~\bibnamefont {{Markham}}}, \bibinfo {author} {\bibfnamefont {D.~J.}\
  \bibnamefont {{Twitchen}}}, \bibinfo {author} {\bibfnamefont
  {D.}~\bibnamefont {{Elkouss}}}, \bibinfo {author} {\bibfnamefont
  {S.}~\bibnamefont {{Wehner}}}, \bibinfo {author} {\bibfnamefont {T.~H.}\
  \bibnamefont {{Taminiau}}}, \ and\ \bibinfo {author} {\bibfnamefont
  {R.}~\bibnamefont {{Hanson}}},\ }\href@noop {} {\bibfield  {journal}
  {\bibinfo  {journal} {ArXiv e-prints}\ } (\bibinfo {year} {2015})},\ \Eprint
  {http://arxiv.org/abs/1508.05949} {arXiv:1508.05949 [quant-ph]} \BibitemShut
  {NoStop}%
\bibitem [{\citenamefont {Clauser}\ \emph {et~al.}(1969)\citenamefont
  {Clauser}, \citenamefont {Horne}, \citenamefont {Shimony},\ and\
  \citenamefont {Holt}}]{ClauserHorneShimonyEtAl-ProposedExperimentto-1969}%
  \BibitemOpen
  \bibfield  {author} {\bibinfo {author} {\bibfnamefont {J.~F.}\ \bibnamefont
  {Clauser}}, \bibinfo {author} {\bibfnamefont {M.~A.}\ \bibnamefont {Horne}},
  \bibinfo {author} {\bibfnamefont {A.}~\bibnamefont {Shimony}}, \ and\
  \bibinfo {author} {\bibfnamefont {R.~A.}\ \bibnamefont {Holt}},\ }\href
  {\doibase 10.1103/PhysRevLett.23.880} {\bibfield  {journal} {\bibinfo
  {journal} {Phys. Rev. Lett.}\ }\textbf {\bibinfo {volume} {23}},\ \bibinfo
  {pages} {880} (\bibinfo {year} {1969})}\BibitemShut {NoStop}%
\bibitem [{\citenamefont {Bavarian}\ and\ \citenamefont
  {Shor}(2015)}]{BavarianShor-InformationCausalitySzemeredi-Trotter-2015}%
  \BibitemOpen
  \bibfield  {author} {\bibinfo {author} {\bibfnamefont {M.}~\bibnamefont
  {Bavarian}}\ and\ \bibinfo {author} {\bibfnamefont {P.~W.}\ \bibnamefont
  {Shor}},\ }in\ \href {\doibase 10.1145/2688073.2688112} {\emph {\bibinfo
  {booktitle} {Proceedings of the 2015 Conference on Innovations in Theoretical
  Computer Science}}},\ \bibinfo {series and number} {ITCS '15}\ (\bibinfo
  {publisher} {ACM},\ \bibinfo {address} {New York, NY, USA},\ \bibinfo {year}
  {2015})\ pp.\ \bibinfo {pages} {123--132}\BibitemShut {NoStop}%
\bibitem [{\citenamefont {Ji}\ \emph {et~al.}(2008)\citenamefont {Ji},
  \citenamefont {Lee}, \citenamefont {Lim}, \citenamefont {Nagata},\ and\
  \citenamefont {Lee}}]{JiLeeLimEtAl-MultisettingBellinequality-2008}%
  \BibitemOpen
  \bibfield  {author} {\bibinfo {author} {\bibfnamefont {S.-W.}\ \bibnamefont
  {Ji}}, \bibinfo {author} {\bibfnamefont {J.}~\bibnamefont {Lee}}, \bibinfo
  {author} {\bibfnamefont {J.}~\bibnamefont {Lim}}, \bibinfo {author}
  {\bibfnamefont {K.}~\bibnamefont {Nagata}}, \ and\ \bibinfo {author}
  {\bibfnamefont {H.-W.}\ \bibnamefont {Lee}},\ }\href {\doibase
  10.1103/PhysRevA.78.052103} {\bibfield  {journal} {\bibinfo  {journal} {Phys.
  Rev. A}\ }\textbf {\bibinfo {volume} {78}},\ \bibinfo {pages} {052103}
  (\bibinfo {year} {2008})}\BibitemShut {NoStop}%
\bibitem [{\citenamefont {Liang}\ \emph {et~al.}(2009)\citenamefont {Liang},
  \citenamefont {Lim},\ and\ \citenamefont
  {Deng}}]{LiangLimDeng-Reexaminationofmultisetting-2009}%
  \BibitemOpen
  \bibfield  {author} {\bibinfo {author} {\bibfnamefont {Y.-C.}\ \bibnamefont
  {Liang}}, \bibinfo {author} {\bibfnamefont {C.-W.}\ \bibnamefont {Lim}}, \
  and\ \bibinfo {author} {\bibfnamefont {D.-L.}\ \bibnamefont {Deng}},\ }\href
  {\doibase 10.1103/PhysRevA.80.052116} {\bibfield  {journal} {\bibinfo
  {journal} {Phys. Rev. A}\ }\textbf {\bibinfo {volume} {80}},\ \bibinfo
  {pages} {052116} (\bibinfo {year} {2009})}\BibitemShut {NoStop}%
\bibitem [{\citenamefont {Dvir}(2010)}]{Dvir-IncidenceTheoremsand-2010}%
  \BibitemOpen
  \bibfield  {author} {\bibinfo {author} {\bibfnamefont {Z.}~\bibnamefont
  {Dvir}},\ }\href {\doibase 10.1561/0400000056} {\bibfield  {journal}
  {\bibinfo  {journal} {Foundations and Trends® in Theoretical Computer
  Science}\ }\textbf {\bibinfo {volume} {6}},\ \bibinfo {pages} {257} (\bibinfo
  {year} {2010})}\BibitemShut {NoStop}%
\end{thebibliography}%

\end{document}